\documentclass[jgr]{AGUTeX}
\usepackage{graphicx}
\usepackage{rotating, lineno, longtable}

\setkeys{Gin}{draft=false}

\authorrunninghead{OLIVEIRA AND RAEDER}

\titlerunninghead{GEOEFFECTIVENESS OF IP SHOCKS}

\authoraddr{Corresponding author: Denny Oliveira, EOS Space Science
  Center, University of New Hampshire, Durham, NH USA. (dmz224@wildcats.unh.edu)}

\begin{document}


\title{Impact angle control of interplanetary shock geoeffectiveness: A statistical study}




\authors{Denny M. Oliveira and Joachim Raeder}

\affil{Department of Physics, and EOS Space Science  Center, University of New Hampshire, Durham, NH USA.}




\begin{abstract}

We present a survey of interplanetary (IP) shocks using WIND and ACE satellite data from January 1995 to December 2013 to study how IP shock geoeffectiveness is controlled by IP shock impact angles. A shock list covering one and a half solar cycle is compiled. The yearly number of IP shocks is found to correlate well with the monthly sunspot number.  We use data from SuperMAG, a large chain with more than 300 geomagnetic stations, to study geoeffectiveness triggered by IP shocks. The SuperMAG SML index, an enhanced version of the familiar AL index, is used in our statistical analysis. The jumps of the SML index triggered by IP shock impacts on the Earth's magnetosphere is investigated in terms of IP shock orientation and speed. We find that, in general, strong (high speed) and almost frontal (small impact angle) shocks are more geoeffective than inclined shocks with low speed. The strongest correlation (correlation coefficient R = 0.70) occurs for fixed IP shock speed and varying the IP shock impact angle. We attribute this result, predicted previously with simulations, to the fact that frontal shocks compress the magnetosphere symmetrically from all sides, which is a favorable condition for the release of magnetic energy stored in the magnetotail, which in turn can produce moderate to strong auroral substorms, which are then observed by ground based magnetometers.

\end{abstract}

\def\RE{R$_\mathrm{E}$ }
\def\thbn{$\theta_{B_n}$}
\def\thxn{$\theta_{x_n}$}
\def\phiyn{$\varphi_{y_n}$}
\def\lam{$\lambda_m$}



\begin{article}

\section{Introduction}

Interplanetary (IP) shocks occur throughout the heliosphere as a result of the interaction of solar disturbances with the solar wind \citep{Burlaga1971a,Richter1985}. As IP shocks interact with the Earth's magnetosphere, they cause disturbances that can be seen throughout the magnetosphere. Some of these disturbances can have implications in several sectors of both magnetosphere and ionosphere \citep{Oliveira2014a}, for example, SSCs (storm sudden commencements), geomagnetic storms, auroral substorms, and GICs (ground-induced currents). GICs may impact power grids, causing electric power disruptions due to equipment damage \citep{Bolduc2002,metatech-report-2010}, and interruption of commercial activities leading to severe economic losses \citep{Schrijver2014}. \par

At 1 AU, most IP shocks are fast shocks. Although slow shocks occur close to the Sun, a few slow shocks may be observed at Earth's orbit \citep{Chao1970,Whang1996}. IP shocks are then classified as forward and reverse. Forward shocks propagate away from the Sun, and reverse shocks propagate towards the Sun in the solar wind frame. Since the solar wind speed is almost always supermagnetosonic, all shocks propagate away from the Sun in the Earth's frame. IP shocks may be further classified by their strength in terms of Mach numbers and the compression ratio, the ratio of  downstream to upstream plasma densities. Among other parameters, the shock normal is another important feature of IP shocks, because shock normal orientations determine how IP shocks propagate throughout the heliosphere. Normals of most IP shocks generated by ICMEs (the interplanetary manifestation of coronal mass ejections at the Sun) at 1 AU are concentrated near the Sun-Earth line \citep{Richter1985}. However, shocks driven by corotating interaction regions (CIRs), as a result of the slow solar wind compression by a fast stream, have normals inclined in relation to the Sun-Earth line \citep[see, e.g.][and references therein]{Siscoe1976,Pizzo1991}. For CIR-driven shocks, the normal angles in the azimuthal direction in relation to the solar coordinate system are generally equal or larger than the inclination angle \citep{Siscoe1976,Pizzo1991}. Calculations of IP shock normals are very sensitive to upstream and downstream plasma parameters, and using data from more than one spacecraft is thought to improve shock normal determinations \citep{Russell2000a}. \par

Here our primary concern is to study the influence of IP shock normal orientations on IP shock geoeffectiveness. Similar studies have been done in the past, but with different space parameters. For example, \cite{Jurac2002} found that the angle between the shock normal and the upstream magnetic field plays an important role in the shock geoeffectiveness. Their main result was the finding that quasi-perpendicular shocks were more geoeffective than quasi-parallel shocks. \cite{Takeuchi2002b} showed that an IP shock highly inclined in relation to the equatorial plane led to an unusual SSC rise time ($\sim$ 30 min). They attributed this result to the fact that the IP shock observed took a longer time to sweep over what they called the ``geoeffective magnetopause''. By performing global MHD simulations, \cite{Guo2005} showed that two similar IP shocks with different shock normal orientations, namely a frontal shock and a highly inclined shock whose shock normals lay in the equatorial plane, interacted with the Earth's magnetosphere leading to different conclusions. They reported that the inclined case took longer than the frontal case to reach fairly similar final quasi-steady states. Later on, \cite{Wang2006}, using ACE and WIND satellite data from 1995 to 2004, reported that, in a survey of nearly 300 fast forward IP shocks, 75\% of them were followed by SSCs observed on the ground. They also found that the shock impact angle plays an important role in determining the SSC rise time, as previously suggested by \cite{Takeuchi2002b}. When the shock speed (shock strength) was fixed, the more parallel the shock normal with the Sun-Earth line, the smaller the rise time. The same occurred when they fixed the shock inclination and changed the shock speed. The faster the shock, the shorter the SSC rise time. \par

More recently, the geoeffectiveness of IP shocks controlled by IP shock impact angles was studied by \cite{Oliveira2014b} through global MHD simulations. Using the OpenGGCM (Open Global Geospace Circulation Model) MHD code \citep{Raeder2003}, they showed that similar IP shocks with different IP shock impact angles may lead to different IP shock geoeffectiveness. They simulated three different IP shocks, where two had shock normals inclined in relation to the Sun-Earth line in the meridian plane, while the second had a Mach number as twice as the first. Both shocks were oblique, i.e., their shock normals were at angles close to 45$^o$ with the upstream magnetic field in the shock frame of reference. Finally, the impact of a third shock, namely a frontal and perpendicular shock with the same Mach number as the first shock, with the Earth's magnetosphere was simulated. \cite{Oliveira2014b} found that the third shock was much more geoeffective than the other two because the shock was frontal, and the magnetosphere was compressed symmetrically on both north and south sides. This compression led then to the triggering of a strong auroral substorm not seen in the other cases. \par

The goal of this work is to confirm the role of IP shock impact angles in the IP shock geoeffectiveness using satellite and geomagnetic activity data. The geomagnetic activity triggered by such interactions is then analyzed using an enhanced version of the geomagnetic AL index. The data and methodology used are discussed in section 2. We present our statistical results in  section 3, and in section 4 we summarize and discuss our results.

\section{Data and methodology}

\subsection{Data}

We screen solar wind plasma and field data at 1 AU to find fast forward IP shock events. In order to do so, we use two different spacecraft close to the equatorial plane: WIND, with data from 1995 up to 2013, and ACE (Advanced Composition Explorer), with data from 1998 also up to 2013. The WIND data were obtained from the Solar Wind Experiment (SWE) instrument \citep{Ogilvie1995}, and the Magnetic Field Investigation (MFI) instrument \citep{Lepping1995}, both in 93-second time resolution. The ACE data were obtained from the Solar Wind Electron Proton Alpha Monitor (SWEPAM) instrument \citep{McComas1998} and the ACE Magnetic Field Experiment (MAG) instrument \citep{Smith1998}, both with 64-second time resolution. All data were downloaded from the CDAWeb interface located at http://cdaweb.gsfc.nasa.gov. All these data were used to compile a list of 461 fast forward IP shock events that can be found in the supplementary material. \par

The monthly averaged sunspot number (SSN) data were compiled by the Solar Influence Data Analysis Center (SIDC). This list was downloaded from http://sidc.oma.be/silso/datafiles. \par

It is well established that substorm activity may be triggered by IP shocks \citep{Kokubun1977,Akasofu1980,Zhou2001,Yue2010,Tsurutani2014b,Oliveira2014b}, and that AL  appears to be the preferred index to quantify the strength of auroral activity \citep{Mayaud1980}. The AE index, the auroral electroject index, was first suggested by \cite{Davis1966} and has been heavily used by magnetospheric physicists since then. However, as pointed out by \cite{Davis1966} themselves and reviewed by \cite{Rostoker1972}, the indices AU, AL, and AE = AU - AL, available at the World Data Center (WDC) in Kyoto, Japan, website (http://wdc.kugi.kyoto-u.ac.jp/aeasy/index.html.), are limited because of the relatively low number of ground stations used to define these indices which happens to be 12. Therefore it is clear that sometimes strong auroral events are underestimated because there are no ground stations under the auroral bulge contributing to the construction of these indices during some strong auroral substorm events \citep{Gjerloev2004}. As an alternative to alleviate this deficiency, SuperMAG, a large worldwide collaboration involving more than 300 ground-based magnetometers, was formed \citep{Gjerloev2009}. Because the AU, AL, and AE indices are recognized as official indices by IAGA, SuperMAG defines SMU as the SuperMAG measurement of the maximum eastward auroral electrojet strength (upper envelope of N-component measured by stations between 40$^o$ and 80$^o$ magnetic north), SML as the SuperMAG measurement of the minimum westward auroral electrojet strength (lower envelope of N-component measured by stations between 40$^o$ and 80$^o$ magnetic north), and SME = SMU - SML as the SuperMAG measurement of the auroral electroject index defined as the distance between the last two indices \citep{Newell2011a}. \par

An example of an auroral substorm event observed by different numbers of IAGA and SuperMAG stations is represented by Figure 1 in \cite{Newell2011a}. In their event, it is shown by Polar UVI imagery that the expansion of the auroral bulge does not pass over any AE ground stations, but rather passes over more than half dozen of the SME ground stations. This auroral substorm was underestimated by the AE stations, as shown by \cite{Newell2011a} in their Figure 2. The AL stations did not detect a substorm event; however, the SML stations recorded a substorm onset 37 seconds after the onset registered by Polar UVI observations. Therefore it is important to mention that AE and SME, besides the other SuperMAG indices, are primarily of the same nature, but with the SuperMAG indices being enhanced by the higher number of ground based stations used to build the SuperMAG indices. More details about the SuperMAG initiative can be found in \cite{Gjerloev2009,Newell2011a}, and an explanation about data techniques and assimilation is reported by \cite{Gjerloev2012}. Finally, the data are available from the SuperMAG websites http://supermag.jhuapl.edu/ and http://supermag.uib.no/.



\subsection{Determination of shock parameters and event analyses}

IP shocks during the period investigated here have been cataloged by several sources, such as the Havard-Smithsonian Center for Astrophysics (CfA) IP shock list compiled by Dr. J. C. Kasper located at http://www.cfa.harvard.edu/shocks/wi$_{-}$data/ for WIND data, and http://www.cfa.harvard.edu/shocks/ac$_{-}$master$_{-}$data/ for ACE data. We also used a shock list compiled by the ACE team available at http://www-ssg.sr.unh.edu/mag/ace/ACElists/obs$_{-}$list.html\#shocks. Another source used was the shock list with only ACE data from February 1998 to August 2008 published by \cite{Wang2010c}. All these lists were merged to compile the shock list used here. We also used an automated search program to detect IP shock candidates in the raw data. After the shock was visually inspected, and if it satisfied the Rankine-Hugoniot conditions, the event was included in our list. Other sources were also consulted for comparison among several events in terms of solar wind conditions and IP shock parameters, such as calculated IP shock normal angles and speeds, when available \citep{Berdichevsky2000, Russell2000a,Zhou2001,Prech2008,Wang2009,Richardson2010b,Koval2010,Zhang2012,Grygorov2014}. \par

Once a shock was identified, solar wind data from WIND and ACE were inspected to provide the basis for IP shock parameter calculations. It is well known that IP shock normal calculations are very sensitive to upstream and downstream plasma parameters. Then, the highest quality available spacecraft data were chosen for shock parameter determinations as described below. From a total of 461 identified fast forward IP shocks, 272 were observed by ACE (59\%), and 189 were observed by WIND (41\%).\par

\begin{figure}[h]
\vspace{-0.0cm}
\hspace*{0.3cm}\includegraphics[width=0.9\hsize]{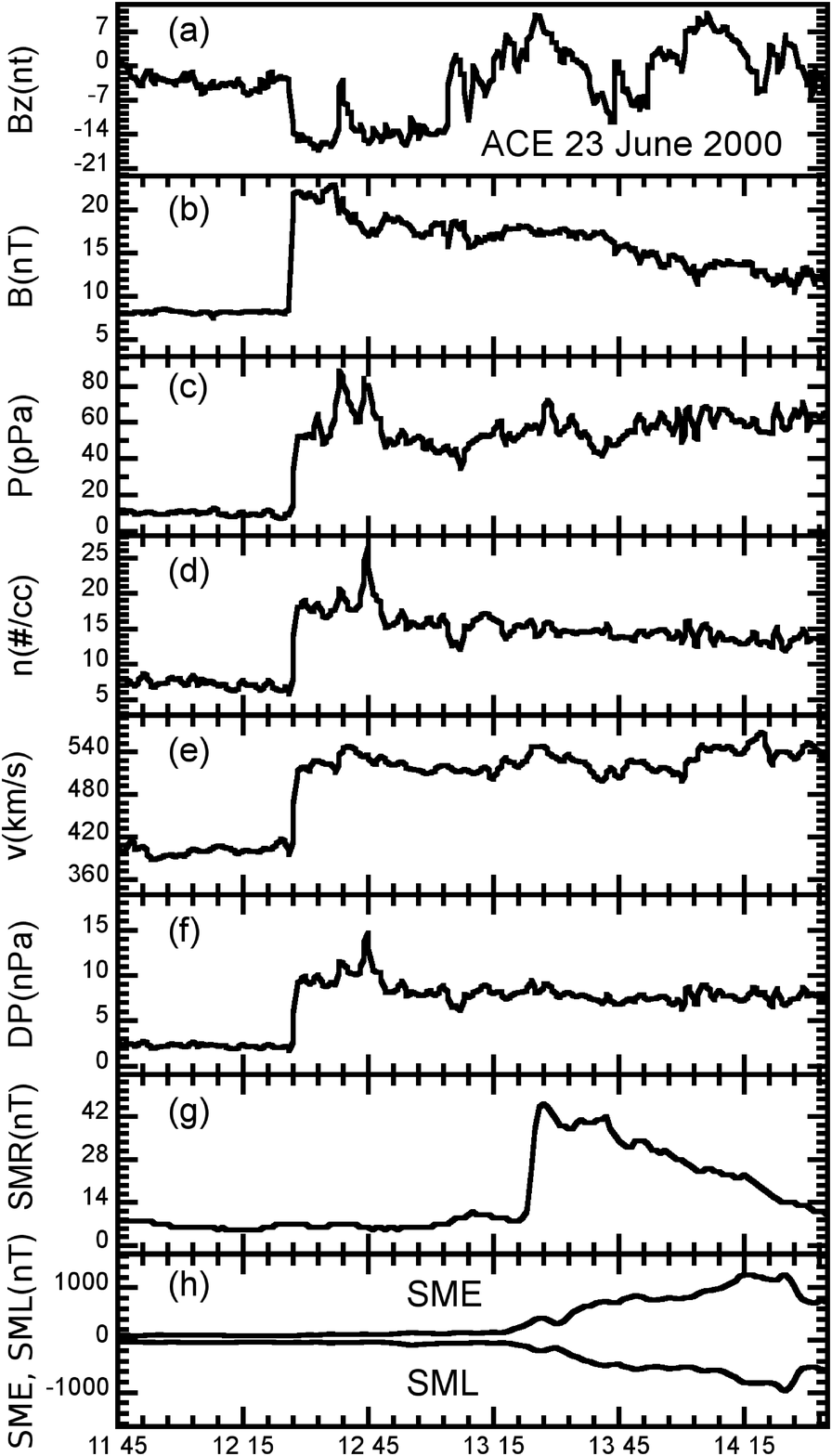}
\vspace{0.1cm}
\caption{The IP shock event seen by ACE on 23 June 2000 at 1227 UT is an example of the methodology used for shock normal calculation and geomagnetic activity analysis. Plots (a)-(f) show jumps in $B_z$ and total magnetic field, in nT; thermal plasma pressure, in pPa; particle number density, in $cm^{-3}$; shock speed, in km/s; and dynamic pressure DP ($\rho v^2$), in nPa. Plots (g) and (h) show SuperMAG data for the symmetric ring current SMR (similar to SYM-H), SME, and SML, all in nT. The maximum geomagnetic activity was recorded for both SME and SML approximately 2 hours after the shock impact. The time interval used to identify peaks in geomagnetic activity for all IP shocks was from $\sim$30 minutes to 2 hours after shock impacts.}
\label{figure_label}
\end{figure}

There have been a variety of shock normal determination methods suggested since late 1960s. Some of the most commonly used methods using single spacecraft data are the magnetic and velocity coplanarity \citep{Colburn1966}, and the mixed IMF and plasma data methods \citep{Abraham-Shrauner1972,Abraham-Shrauner1976,Vinas1986}. Although situations where data from more than one spacecraft are available give more reliable results \citep{Burlaga1980,Russell1983,Russell1984a,Russell1984b,Thomsen1988,Russell2000a,Szabo2005,Koval2010}, we use the methods based only on one spacecraft. Multiple spacecraft data usage would create an inconsistent data set in a large statistical study, because the availability of more than one spacecraft data for shock normal determination is rare. The IP shock normals are then calculated using the methods of magnetic coplanarity (MC), velocity coplanarity (VC), and the three mixed data methods (MX1,MX2,MX3) found in  \cite{Schwartz1998}. Then, the average of the at least three closest results is calculated and registered as the chosen IP shock normal for each event. \par

An example of an event analysis is shown in Figure ~1. This shock event was seen by ACE on 23 June 2000. At 1227 UT and (234, 36.6,-0.7)\RE GSE upstream of the Earth, ACE observed sharp jumps in magnetic field $B_z$ component, total magnetic field, plasma thermal pressure, particle number density, plasma velocity, and dynamic pressure $\rho v^2$ (Figure~ 1(a)-(f)). Approximately 55 minutes later, the shock impacted the magnetopause, the magnetosphere was compressed by the shock and an SSC was detected by SuperMAG geomagnetic stations, as can be seen in Figure 1(g) for the SuperMAG symmetric ring current index SMR \citep{Newell2012}, the SuperMAG index similar to the well known SYM-H index. Increases in the SuperMAG indices SME and SML followed the IP shock approximately 2 hours after shock impact, reaching a maximum of about 1500 nT for SME and a minimum of about -1000 nT for SML.  The calculated shock normal of this event is (-0.785,0.153,-0.600), with $\theta_{x_n}\sim$140$^o$, shock speed of 553.2 km/s, and fast magnetosonic Mach number 2.60. Using these results, and assuming the estimated position of the magnetopause previous to the shock impact as 10 $R_E$ as suggested by \cite{Zhou1999}, the calculated time travel is $\sim$55 minutes, in agreement with observations, which validates our method. To complete the shock property analysis, the compression ratio (the ratio of downstream to upstream plasma density) was 2.62, and the fast magnetosonic Mach number was 2.60. 

\begin{figure}
\vspace{0.08cm}
\hspace*{0.0cm}\includegraphics[width=1.0\hsize]{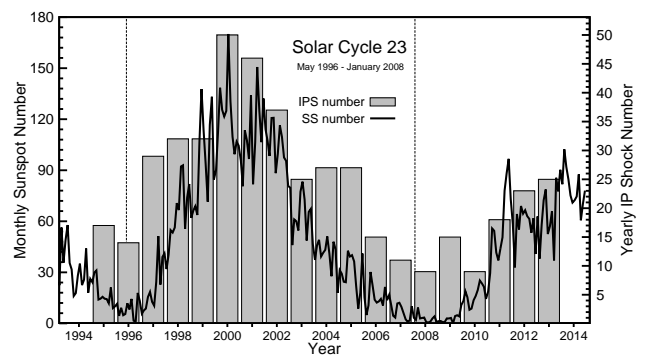}
\vspace{0.1cm}
\caption{Yearly IP shock number (gray bars) plotted against the SIDC monthly sunspot number (solid line). WIND and ACE data were used to identify all IP shock events. A strong correlation can be seen. The maximum yearly IP shock number occurred in the year 2000 (50 events), in the solar maximum of the solar cycle 23. Due to the unusually low sunspot number in the maximum of the current solar cycle, only 25 events were observed in 2013, and not many more are expected to be identified in the 2014 and 2015 WIND and ACE data \citep{Smith2014}.}
\label{figure_label}
\end{figure}

\section{Statistical results}

\subsection{Solar wind and shock parameters}

Figure 2 shows the yearly IP shock number (gray bars) and the monthly sunspot number (SSN, solid lines) plotted in the time range from 1995 to 2013. This time period includes the whole solar cycle 23, which ranged from May 1996 to January 2008. A correlation between both numbers is clear. During the ascending phase of the solar cycle 23, the number of IP shocks increased with the SSN. Correlations between the number of SSCs and the SSN in different solar cycles have been reported by earlier works \citep{Chao1974,Hundhaunsen1979,Smith1983,Smith1986,Rastogi1999}. Since 75\% of SSCs are caused by IP shocks \citep{Wang2006}, these arguments are considered to be very similar. Such correlations may be a result of the increasing number of shocks driven by ICMEs \citep{Jian2006b}, since the number of CIR-driven shocks did not show any correlation with the solar cycle phase \citep{Jian2006a}. Then, during the declining phase of the solar cycle 23, the number of IP shocks decreases with the SSN, since the number of ICMEs related to shocks decreases. The number of fast forward shocks is higher in the maximum of the solar cycle 23  (year 2000, 50 events) than in the solar minimum of the same solar cycle (years 1995-1996, 31 events) \citep{Echer2003a}. Due to the unusual low SSN of the current solar cycle maximum, barely more than 25 shocks are expected to be found in the WIND and ACE data for 2014 and even 2015 \citep{Smith2014}. \par

\begin{figure}
\vspace{0.05cm}
\hspace*{-0.2cm}\includegraphics[width=0.9\hsize]{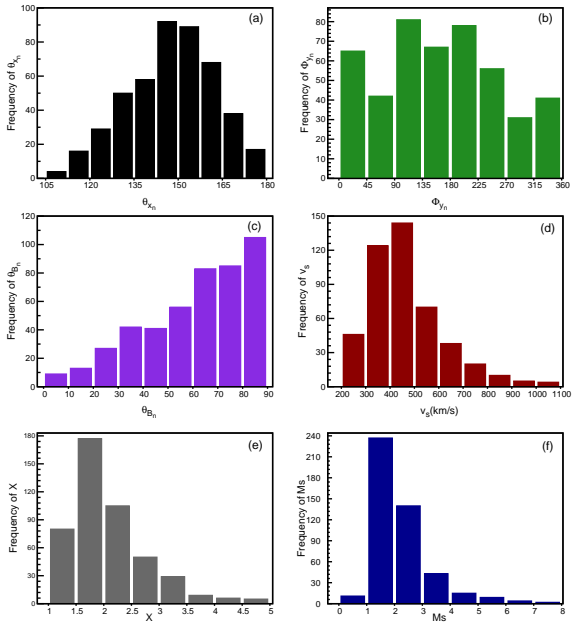}
\vspace{0.1cm}
\caption{Statistical results of the full list with 461 IP shocks. Figure 3(a) shows the distribution of the impact angle between the shock normal and the Sun-Earth line. Angles close to 180$^o$ represent almost frontal shocks. The clock angle $\varphi_{y_n}$ distribution on the GSE YZ plane is shown in Figure 3(b). Angles in the ranges 0$^o$ $\leq\, \varphi_{y_n}\,\leq$ 45$^o$, 135$^o$ $\leq\, \varphi_{y_n}\,\leq$ 225$^o$, and 315$^o$ $\leq\, \varphi_{y_n}\,\leq$ 360$^o$ indicate that the shock normals were close to the equatorial plane. Figure 3(c) shows the distribution of $\theta_{B_n}$, the angle between the upstream magnetic field vector and the shock normal. Represented in Figure 3(d), is the shock speed (in km/s) distribution, in relation to the spacecraft frame of reference. Figure 3(e) shows the distribution of the compression ratio, the ratio of the downstream to upstream plasma densities. Finally, the fast magnetosonic Mach number distribution is shown by Figure 3(f).}
\label{figure_label}
\end{figure}

A statistical analysis of solar wind and IP shock parameters is shown in Figure 3(a-f). Figure 3(a) shows $\theta_{x_n}$, the angle between the shock normal vector and the Sun-Earth line. Angles close to 180$^o$ indicate IP shocks whose shock normals were almost parallel to the Sun-Earth line, or almost frontal shocks. IP shocks with angles close to 90$^o$ represent inclined shocks. In our list, 363 (78.57\%) cases had shocks with  $\theta_{x_n}\ge135^o$. The distribution of the clock angle $\varphi_{y_n}$ is shown in Figure 3(b). Shock normals with 0$^o$ $\leq\, \varphi_{y_n}\,\leq$ 45$^o$, 135$^o$ $\leq\, \varphi_{y_n}\,\leq$ 225$^o$, and 315$^o$ $\leq\, \varphi_{y_n}\,\leq$ 360$^o$ indicate that the shock normal was close to the equatorial plane. These conditions were satisfied by 276 events, or 59.74\%. Figure 3(c) shows the obliquity $\theta_{B_n}$, the angle between the shock normal and the upstream magnetic field vector. In our data set,  354 cases showed $\theta_{B_n}$ larger than 45$^o$, and most of the shocks in this category might have been driven by ICMEs \citep{Richardson2010b}. The shock speed distribution is shown in Figure 3(d). The average shock speed is 467 km/s, and it tends to be higher in solar maximum, and lower in solar minimum, as already reported by \cite{Berdichevsky2000} and \cite{Echer2003a} with data partially in the same time period. The percentage of shocks above the average speed is 40.13\%, or 185 events. The compression ratio, the ratio of the downstream to upstream plasma densities, can be seen in Figure 3(e). As reported before \citep{Berdichevsky2000}, most shocks have their compression ratios between 1.2 and 2.0, which happened to 251 of our cases (54.44\%). Our compression ratio average is 2.07. Although the theoretical limit for the compression ratio is 4 \citep{Richter1985}, which is derived for perpendicular shocks,  this value was exceeded in 11 cases (2.38\%), and most of them took place slightly before and after the  solar maximum (year 2000). \cite{Echer2003a} argued that such cases can happen for some shocks in a data set in which shock obliquities range from almost parallel to almost perpendicular shocks. Finally, the fast magnetosonic Mach number distribution is shown in Figure 3(f). The average of Ms is 2.15, and it is clear that most shocks have Ms between 1.0 and 2.0 \citep{Kallenrode2003}. The number of shocks with Ms above the average is 166 (36.00\%). However, some shocks have Ms less than one, which can be an indication that such events were not shocks because the shock waves could not steepen, even though they could show some shock-like behavior \citep{Kennel1985}. These events were not included in our statistical analysis. Therefore, as a consequence of this analysis, it is possible to conclude that the interplanetary space is dominated by weak IP shocks. The agreement of our results with other works validates our statistical analysis, in particular the shock normal determination methods used in this work.\par

\subsection{Geomagnetic activity}

In this section we investigate the geoeffectiveness of IP shocks by correlating the shock parameters with the SuperMAG SML index as a geomagnetic activity indicator. Changes in this index, $\Delta$SML, in nT, are recorded for each event from $\sim$30 minutes to two hours after shock impact. If the IP shock is followed by any other solar wind structure, only the first peak in the data is considered. We chose this time frame because some inclined shocks take a long time to sweep over the magnetosphere when they are inclined in relation to the Sun-Earth line \citep{Takeuchi2002b,Guo2005,Wang2006,Oliveira2014b}. We used SuperMAG data up to 2013 because the 2014 SuperMAG data were not yet available.\par

\begin{figure}
\vspace{0.08cm}
\hspace*{-0.1cm}\includegraphics[width=0.9\hsize]{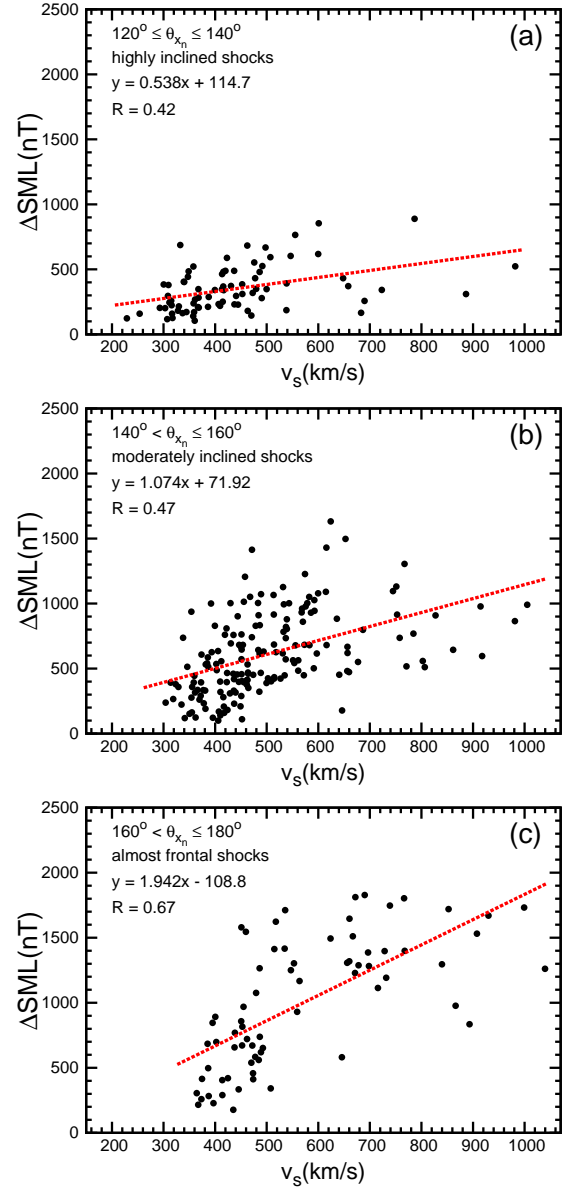}
\vspace{0.1cm}
\caption{SML jumps, in nT, triggered by IP shock impacts are plotted as a function of the shock speed $v_s$, in km/s. The events were binned in three different groups in terms of the shock orientation in relation to the Sun-Earth line: Figure 4(a), 120$^o$ $\leq\,\theta_{x_n}\,\leq$ 140$^o$ (highly inclined shocks); Figure 4(b), 140$^o$ $<\,\theta_{x_n}\,\leq$ 160$^o$ (inclined shocks); and Figure 4(c) 160$^o$ $<\,\theta_{x_n}\,\leq$ 180$^o$ (almost frontal shocks). The IP shocks are more geoeffective for strong (high speed) and almost frontal shocks (large $\theta_{x_n}$).}
\label{figure_label}
\end{figure}

Figure 4 shows jumps in SML, in nT, measured by SuperMAG ground stations plotted against the shock speed, in km/s. Since we consider two parameters, shock speed and impact angle, all the data were binned in three different groups in terms of the shock normal impact angle $\theta_{x_n}$. Here, the impact angle is held and the shock speed varies. Figure 4(a) shows highly inclined shocks: 120$^o$ $\leq\,\theta_{x_n}\,\leq$ 140$^o$; Figure 4(b) represents moderately inclined shocks, 140$^o$ $<\,\theta_{x_n}\,\leq$ 160$^o$; and almost frontal shocks, 160$^o$ $<\,\theta_{x_n}\,\leq$ 180$^o$, can be found in Figure 4(c). In Figure 4(a), most shocks  produce little geomagnetic activity ($\Delta$SML $<$ 500 nT), and in such cases most shocks had $v_s\,<$ 450 km/s. This is expected for weak and highly inclined shocks. For some stronger, but highly inclined shocks, the resulting activity is slightly larger, but just a few such shocks in this case were identified in the data. The linear regression analysis gives a correlation coefficient of R = 0.34. In the intermediate case, i.e., the case of shocks with moderate inclination, most shocks produced $\Delta$SML $>$ 500 nT. In this case, there is a stronger correlation. We attribute the correlation coefficient of R = 0.41 to the fact that most shocks with $v_s\,<$ 450 km/s triggered small jumps in SML ($\Delta$SML $<$ 500 nT). For the cases in which $v_s\,>$ 450 km/s, $\Delta$SML showed better correlations, but just a few with $\Delta$SML $>$ 1000 nT. In the more extreme case, namely the case in which the IP shocks were almost frontal, the correlation coefficient is R = 0.60. In this case, approximately half of the shocks with $v_s\,<$ 450 km/s did not show large jumps in SML. Most shocks triggered $\Delta$SML $>$ 500 nT, and almost all cases in which $\Delta$SML $>$ 1000 nT had $v_s$ larger than 450 km/s. Therefore, by inspecting all plots, it is clear that the IP shock geoeffectiveness increases with both shock strength and shock impact angle. Table 1 summarizes the results obtained in all categories in this case. \par

\begin{figure}
\vspace{0.08cm}
\hspace*{-0.1cm}\includegraphics[width=0.9\hsize]{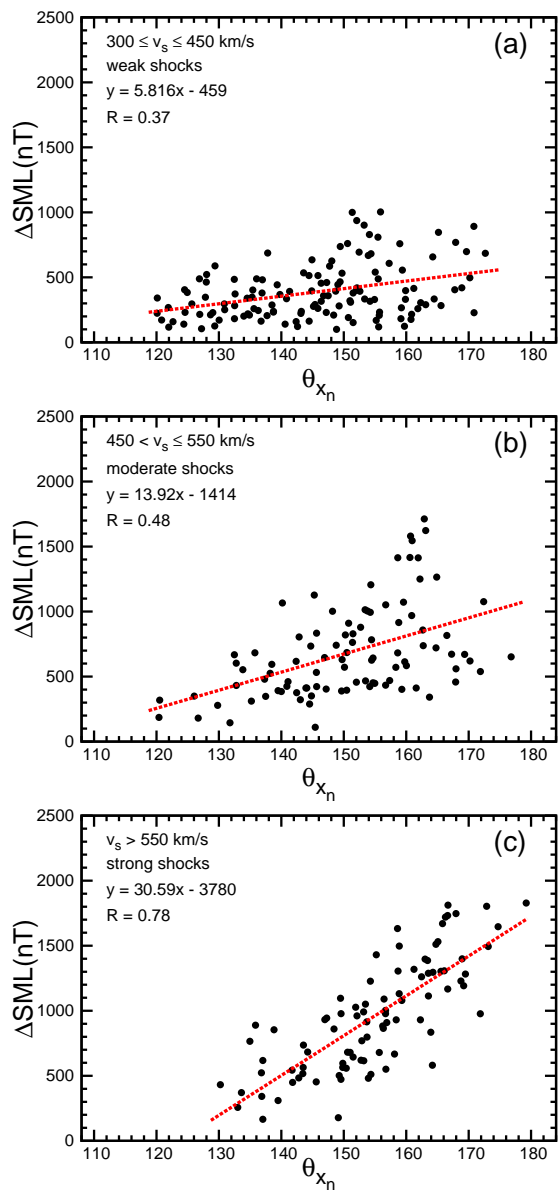}
\vspace{0.1cm}
\caption{SML jumps, in nT, triggered by IP shock impacts are plotted as a function of the shock impact angle $\theta_{x_n}$, in km/s. The events were binned in three different groups in terms of the shock speed: Figure 5(a), 350 $\leq\,v_s\,\leq$ 450 km/s (weak shocks); Figure 5(b), 450 $<\,v_s\,\leq$ 550 km/s (moderate shocks); and Figure 5(c) $v_s\,>$ 550 km/s (strong shocks). The shocks are more geoeffective for almost frontal (large $\theta_{x_n}$) and strong (high speed) IP shocks.}
\label{figure_label}
\end{figure}

The opposite analysis is shown in Figure 5, i.e., the shock speed is held and the impact angle varies. There, $\Delta$SML is plotted against $\theta_{x_n}$, and the data are binned in three different categories related to the shock strength (or shock speed). Figure 5(a) shows the weak shocks, 300 $\leq\,v_s\leq$ 450 km/s; 5(b) moderate shocks, 450 $<\,v_s\leq$ 550 km/s; and 5(c) strong shocks, $v_s\,>$ 550 km/s. Figure 5(a) shows the largest number of small $\Delta$SML ($\Delta$SML $<$ 500 nT), even for shocks with shock normals almost parallel to the Sun-Earth line. The correlation coefficient in this case is R = 0.36. A clearer $\Delta$SML-$\theta_{x_n}$ correlation is evident in the intermediate case, where R = 0.51, and most shock events have $\Delta$SML $>$ 500 nT and $\theta_{x_n}\,>\,135^o$. All shocks with $\Delta$SML $>$ 1000 nT had impact angles larger than 140$^o$. In the category of strong shocks, only a few shocks triggered geomagnetic activity with $\Delta$SML $<$ 500 nT, most of them being  highly inclined shocks in which $\theta_{x_n}\,<\,150^o$. Shocks with high geoeffectiveness, or $\Delta$SML $>$ 1000 nT, were almost frontal shocks with $\theta_{x_n}\,>\,150^o$ (only one event had $\theta_{x_n}$ slightly less than $150^o$ in this case). The highest correlation coefficient, R = 0.70, occurs for IP shocks in this category. Table 1 summarizes the results obtained in all cases in this correlation analysis.\par

Thus, strong shocks are generally much more geoeffective than weak shocks, and the geoeffectiveness increases if the IP shock impacts more frontally the Earth's magnetosphere. These results have already been shown by \cite{Wang2006} for the SSC rise-time and \cite{Oliveira2013,Oliveira2014b} in global MHD simulations.

\begin{table}
\begin{tabular}{c c c c}
\hline
\hline
\multicolumn{4}{c}{{\bf Fixed impact angle $\theta_{x_n}$, changed shock speed $v_s$}} \\
\hline
category & highly inclined & moderately inclined & almost frontal \\
R & 0.42 & 0.47 & 0.67 \\
\hline
\hline
\multicolumn{4}{c}{{\bf Fixed shock speed $v_s$, changed impact angle $\theta_{x_n}$}} \\
\hline
 category & weak & moderate & strong \\
R & 0.37 & 0.48 & 0.78 \\
\hline
\end{tabular}
\caption{Summary of the results obtained for the shock speed, shock impact angle, and $\Delta$SML correlation analyses.}
\end{table}

\section{Summary and Conclusions}

We investigated WIND and ACE solar wind data at 1 AU to compile a list of fast forward interplanetary (IP) shocks. We studied the geoeffectiveness triggered by the IP shock impacts, as measured by the jumps in the SuperMAG SML index, and how it relates to the shock speed (strength) and the shock inclination angles. Our main results are summarized below:

\begin{enumerate}
\item We provide the community with a fast forward IP shock list with events from January 1995 to December 2013, covering the whole solar cycle 23 and half of the current solar cycle.

\item The number of yearly IP shocks correlates closely with the monthly sunspot number. The highest number of fast forward IP shocks was found in the year 2000, the solar maximum of the solar cycle 23. As expected, the number of IP shocks is smaller in the maximum of the current solar cycle due to the unusual low number of sunspots occurring in this period.

\item The majority of the events (76\%) are almost perpendicular
  shocks, with $\theta_{B_n}\geq45^o$. Most shocks (78\%) have their shock normals close to the
    Sun-Earth line, or $\theta_{x_n}\geq135^o$. Also, less than half of the shocks (40\%) have their speeds above the average of about 450 km/s, and shocks with the supermagnetosonic Mach number greater than the average 2.1 was 36\%. These results indicate that the heliosphere at 1 AU is dominated by weak interplanetary shocks.

\item Strong (high speed) shocks are more geoeffective than weak shocks (low speed). The correlation is clearer when shocks are grouped in categories related to their strength and then investigated in terms of their shock impact angles. The largest correlation occurs (R = 0.70) when we fixed the IP shock speed and changed the IP shock impact angles. Thus, the IP shock impact angle is just as important determining their geoeffectiveness as their strength.  This result was predicted by \cite{Oliveira2014b}.
\end{enumerate}

\begin{acknowledgments}
This work was supported by grant AGS-1143895 from the National Science Foundation and grant FA-9550-120264 from the Air Force Office of Sponsored Research. We thank the WIND and ACE teams for the solar wind data and CDAWeb interface for data availability. We thank Dr. C. W. Smith, the ACE team, and Dr. J. C. Kasper for their list compilations. For the ground magnetometer data we gratefully acknowledge: Intermagnet; USGS, Jeffrey J. Love; CARISMA, PI Ian Mann; CANMOS; The S-RAMP Database, PI K. Yumoto and Dr. K. Shiokawa; The SPIDR database; AARI, PI Oleg Troshichev; The MACCS program, PI M. Engebretson, Geomagnetism Unit of the Geological Survey of Canada; GIMA; MEASURE, UCLA IGPP and Florida Institute of Technology; SAMBA, PI Eftyhia Zesta; 210 Chain, PI K. Yumoto; SAMNET, PI Farideh Honary; The institutes who maintain the IMAGE magnetometer array, PI Eija Tanskanen; PENGUIN; AUTUMN, PI Martin Conners; DTU Space, PI Dr. J\"urgen Matzka; South Pole and McMurdo Magnetometer, PI's Louis J. Lanzarotti and Alan T. Weatherwax; ICESTAR; RAPIDMAG; PENGUIn; British Artarctic Survey; McMac, PI Dr. Peter Chi; BGS, PI Dr. Susan Macmillan; Pushkov Institute of Terrestrial Magnetism, Ionosphere and Radio Wave Propagation (IZMIRAN); GFZ, PI Dr. J\"urgen Matzka; MFGI, PI B. Heilig; IGFPAS, PI J. Reda; University of L’Aquila, PI M. Vellante; SuperMAG, PI Jesper W. Gjerloev. D.M.O. thanks the SuperMAG PI J. W. Gjerloev for the straightforward SuperMAG website and its convenience of data visualization and download. The data used in this work can be obtained from the corresponding author at dennymauricio@gmail.com
\end{acknowledgments}


\end{article}

\end{document}